\documentclass[9pt, conference]{IEEEtran}
\IEEEoverridecommandlockouts
\usepackage{cite}
\usepackage{amsmath,amssymb,amsfonts}
\usepackage{algorithmic}
\usepackage{graphicx}
\usepackage{textcomp}
\usepackage{xcolor}
\def\BibTeX{{\rm B\kern-.05em{\sc i\kern-.025em b}\kern-.08em
    T\kern-.1667em\lower.7ex\hbox{E}\kern-.125emX}}
\usepackage{booktabs}
\usepackage{multirow}
\usepackage{multicol}
\usepackage{url}
\usepackage{caption}
\usepackage{subcaption}
\newcommand{\linebreakand}{%
  \end{@IEEEauthorhalign}
  \hfill\mbox{}\par
  \mbox{}\hfill\begin{@IEEEauthorhalign}
}
\begin{document}

\title{AccentBox: Towards High-Fidelity Zero-Shot Accent Generation\\
}
\author{
\IEEEauthorblockN{Jinzuomu Zhong$^1$, Korin Richmond$^1$, Zhiba Su$^2$, Siqi Sun$^1$}
\IEEEauthorblockA{\textit{$^1$Centre for Speech Technology Research, University of Edinburgh, UK}\\
\textit{$^2$Department of AI Technology, Transsion, China}
\\mailto: jinzuomu.zhong@ed.ac.uk, korin.richmond@ed.ac.uk}
}
\maketitle

\fontsize{9}{11}\selectfont
\begin{abstract}
While recent Zero-Shot Text-to-Speech (ZS-TTS) models have achieved high naturalness and speaker similarity, they fall short in accent fidelity and control. To address this issue, we propose zero-shot accent generation that unifies Foreign Accent Conversion (FAC), accented TTS, and ZS-TTS, with a novel two-stage pipeline. In the first stage, we achieve state-of-the-art (SOTA) on Accent Identification (AID) with 0.56 f1 score on unseen speakers. In the second stage, we condition a ZS-TTS system on the pretrained speaker-agnostic accent embeddings extracted by the AID model. The proposed system achieves higher accent fidelity on inherent/cross accent generation, and enables unseen accent generation.
\end{abstract}

\begin{IEEEkeywords}
Accent Generation, Zero-Shot TTS, Accent Identification
\end{IEEEkeywords}
\vspace{-1em}
\section{Introduction}

\subsection{Motivation: Accent Matters in ZS-TTS}
\label{ssec:intro_motivation}
\vspace{-0.25em}
Recent advances in Zero-Shot Text-to-Speech (ZS-TTS) have enabled speech generation of any unseen speaker's voice in a 3-second audio clip, that is on-par quality with human recordings \cite{ju2024naturalspeech, chen2024vall}. However, most ZS-TTS systems focus on replicating speakers' voices \cite{NEURIPS2018_6832a7b2} while largely ignoring accent variation, training on mostly American English data without accent conditioning or control. Such disregard for accents and biased training leads to poor accent fidelity and no control over accents in the generated speech \cite{wang2023neural}.

For native speakers (L1), having their accents accurately generated preserves their linguistic identity, integral to their personal and regional identity \cite{rosina1997english}. For non-native speakers (L2), TTS systems that retain L2 accents can alleviate the pressure to conform to native accents \cite{gluszek2010way}, while enhancing personalized language learning through Computer-Aided Pronunciation Training (CAPT) systems \cite{felps2009foreign, agarwal2019review}.

Motivated by the poor accent generation in ZS-TTS as well as the social and moral imperative for inclusive speech technology, we take an initiative to address accent-related issues in ZS-TTS. Generating accented speech in a zero-shot manner has broad and promising applications in personalised virtual assistants \cite{pal2019user}, movie dubbing \cite{spiteri2021exploring}, CAPT \cite{felps2009foreign, agarwal2019review}, and so on.

\subsection{Task Definition: Zero-shot Accent Generation}
\label{ssec:intro_task}

\begin{table}[h!]
\vspace{-1em}
\centering
\setlength\tabcolsep{1.6pt}
\caption{Different tasks proposed for generating accented speech.}
\label{tab:tts_tasks}
\vspace{-0.5em}
\begin{tabular}{cccc}
\toprule
\multirow{2}{*}{Task} & \multicolumn{3}{c}{Accent Generation Abilities} \\ \cmidrule{2-4} 
 & Any given text? & Any given speaker? & Any given accent? \\ \midrule
\renewcommand*{\arraystretch}{0.7}\begin{tabular}[c]{@{}c@{}}Foreign Accent\\Conversion (FAC)\end{tabular} &
  No. &
  Yes. &
  \renewcommand*{\arraystretch}{0.7}\begin{tabular}[c]{@{}c@{}}Only seen/trained\\ accent pairs.\end{tabular} \\[5pt]
\renewcommand*{\arraystretch}{0.7}\begin{tabular}[c]{@{}c@{}}Multi-Accent/\\ Accented TTS\end{tabular} &
  Yes. &
  \renewcommand*{\arraystretch}{0.7}\begin{tabular}[c]{@{}c@{}}Only seen\\ speakers.\end{tabular} &
  \renewcommand*{\arraystretch}{0.7}\begin{tabular}[c]{@{}c@{}}Only seen\\ accents.\end{tabular} \\[5pt]
Zero-Shot TTS                                                         & Yes.            & Yes.               & No.               \\[5pt]
\renewcommand*{\arraystretch}{0.7}\begin{tabular}[c]{@{}c@{}}Zero-Shot\\ Accent Generation\end{tabular} & Yes.            & Yes.               & Yes.              \\ \bottomrule
\end{tabular}
\end{table}

\vspace{-0.5em}
Previous studies on generating accented speech can be categorised into three related tasks: \textit{1) Foreign Accent Conversion (FAC)} is a speech-to-speech task that takes source speech from a target speaker as input, and converts the L2 accent in the source speech to a target L1 accent \cite{zhao19f_interspeech}. 
However, FAC cannot generate accented speech for any given text or generalise to unseen accent pairs. \textit{2) Accented TTS} aims to generate accented speech with high naturalness and accent fidelity with target text, accent ID, and speaker ID as input, leveraging multi-accent front-ends \cite{10439064, accentvits}, Variational Auto-Encoder (VAE) \cite{10023072}, Diffusion \cite{deja23_interspeech}, phoneme- and utterance-level representation learning \cite{zhou2024multi, liu23u_interspeech, 10487819}. 
Despite these studies, accented TTS remains limited by its inability to generate speech for unseen speakers or unseen accents. \textit{3) ZS-TTS} generates speech using the voice in a speech prompt (i.e.\ reference speech) and target text as input. Voice information derives from either speaker embeddings extracted by a pretrained speaker verification model \cite{NEURIPS2018_6832a7b2, pmlr-v162-casanova22a} or audio/speech codecs in Large Language Modelling (LLM)-based TTS \cite{wang2023neural, kharitonov2023speak, NEURIPS2023_2d8911db}. However, none of these studies adequately addresses accent generation, with some acknowledging poor ZS-TTS performance for accented speakers \cite{wang2023neural}.

We propose a new task: \textbf{Zero-Shot Accent Generation}, which generates any speech content in any given voice and accent from one audio clip, unifying the capabilities of all three tasks mentioned above (see Tab.\ \ref{tab:tts_tasks}). 

\subsection{Research Gap: Speaker-Accent Entanglement in AID and ZS-TTS}
\vspace{-0.25em}

Ideally, a speech dataset should include utterances from the same speaker in different accents. However, most speakers cannot consistently produce a wide range of accents, leading to speaker-accent entanglement issues in both AID and ZS-TTS models.

In AID, the AESRC2020 benchmark \cite{9413386} has been a standard. However, this data is no longer openly available. A more recent benchmark, CommonAccent \cite{zuluagagomez23_interspeech}, uses a subset of Common Voice \cite{ardila-etal-2020-common}, which is open-source and representative of in-the-world speech data. However, our examination of the processing scripts reveals an overlap of speakers across training/validation/testing sets. The extent to which speaker-accent entanglement impacts AID performance remains unexplored, particularly when no effort is made to separate unseen speakers for testing.

In ZS-TTS, the closest to our work is Zhang et al.\ \cite{zhang2023towards,10317526}. They adapt a pretrained Tacotron 2-based \cite{8461368} ZS-TTS, with accent ID as input and AID as auxiliary training objective, to perform zero-shot generation for seen accents. Apart from the limitations of accented TTS, their work: 1) uses limited TTS data to learn accent embeddings, 2) relies on pre-collected accent labels in TTS data, and 3) lacks disentanglement between accent and speaker. Another closely related work by Lyth and King \cite{lyth2024natural} trains an AID model to pseudo-label the data and then uses pseudo-generated text descriptions of the speech to control different attributes (incl.\ accent) in text-guided ZS-TTS. However, their work is: 1) close-sourced, with no accent generation in its open-source reproduction, Parler-TTS\footnote{\url{https://github.com/huggingface/parler-tts}}; 2) unclear about how the AID is trained, and susceptible to speaker-accent entanglement; 3) disregarding the continuous nature of accents with pseudo-labelled discrete accent labels as TTS input condition; and 4) unable to disentangle and separately control speaker and accent in speech generation.

To address these limitations, we first propose to obtain pretrained accent embeddings from an improved AID model with speaker-accent disentanglement, termed generalisable accent identification across speakers (GenAID). This approach offers several benefits: 1) leveraging more non-TTS data to cover more speakers and accents, 2) treating accents as continuous with varying embeddings across different utterances and speakers of the same accent label, and 3) achieving greater generalisability across speakers. We then propose to condition a pretrained YourTTS-based \cite{pmlr-v162-casanova22a} ZS-TTS on these pretrained accent embeddings, named AccentBox. AccentBox is capable of high-fidelity zero-shot accent generation and offers several advantages: 1) leveraging continuous, speaker-agnostic GenAID embeddings, 2) capable of generating unseen accents, 3) no reliance on pre-collected accent labels in TTS data, and 4) providing separate control over speaker and accent in speech generation. Readers are encouraged to visit our demo page\footnote{\url{https://jzmzhong.github.io/AccentBox-ICASSP2025/}} where we include audio samples for accent mismatch/hallucination in current SOTA ZS-TTS (part I) and comparison between different systems and the proposed AccentBox (part IV).
To summarise, our contributions are:
\begin{itemize}
    \item 
    To the best of our knowledge, we are the first to 1) verify and quantify the \textit{speaker-accent entanglement} issue in AID data/models, and 2) highlight the \textit{accent mismatch/hallucination} issue in ZS-TTS.
    \item 
    We introduce novel speaker-accent disentanglement with information bottleneck and adversarial training in AID. We propose the zero-shot accent generation task and set the first benchmark for such task, unifing FAC, accented TTS, and ZS-TTS.
    \item 
    We achieve SOTA results in both AID (0.56 f1 score on unseen speakers in 13-accent classification by GenAID) and zero-shot accent generation (57.4\%-70.0\% accent similarity preference across inherent/cross accent generation against strong baselines by AccentBox).
\end{itemize}
\vspace{-0.5em}
\section{Method}

\subsection{GenAID: Generalisable Accent Identification Across Speakers}
\label{ssec:genaid}

\begin{figure}[h!]
\vspace{-1.5em}
\centering
\includegraphics[width=0.75\linewidth, height=0.5\linewidth, trim = 20 590 220 30, clip]{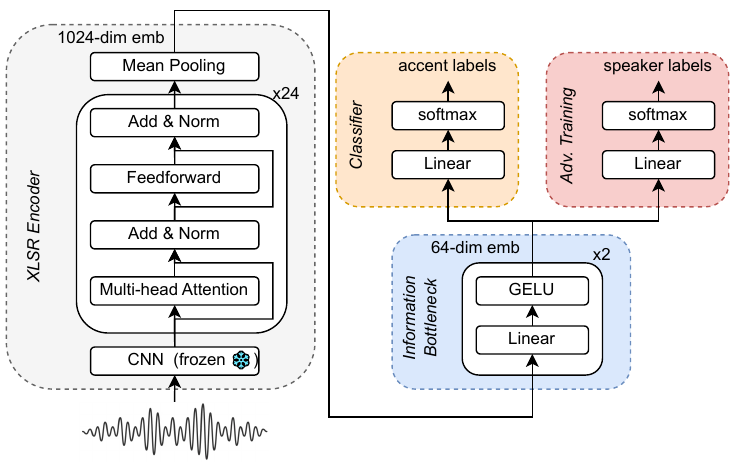}
\captionsetup{justification=centering}
\caption{Model architecture of GenAID.
}
\label{fig:aid_methods}
\vspace{-1em}
\end{figure}

In the first stage, we aim to extract continuous and speaker-agnostic accent embeddings to represent varying accents in speech. We propose an AID model that generalises across speakers, denoted GenAID (see Fig.\ \ref{fig:aid_methods}). Building upon CommonAccent \cite{zuluagagomez23_interspeech} which finetunes XLSR \cite{babu22_interspeech} for AID, we propose five modifications:

\subsubsection{Validation on Unseen Speakers}
\label{sssec:aid_method_1}

To prevent the model from overfitting to seen speakers (by memorising the speaker-to-accent mapping without learning to discriminate accents), we reprocess the data and validate the model only on unseen speakers.

\subsubsection{Weighted Sampling}
\label{sssec:aid_method_2}

To handle imbalanced distribution of accent labels, we apply weighted sampling, to ensure equal probability of sampling each accent's data in each batch \cite{ling1998data}. The sampling weights are the inverse frequency of each accent in the data.

\subsubsection{Data Augmentation by Perturbation}
\label{sssec:aid_method_3}

To make the model more agnostic to various speech factors (e.g.\ recording device, recording environment, speaking rate, etc.), we augment the data by conducting speed \cite{ko15_interspeech} and noise perturbation \cite{7953152}, 
same as CommonAccent \cite{zuluagagomez23_interspeech}.

\subsubsection{Information Bottleneck}
\label{sssec:aid_method_4}

To remove redundant information, especially from the pretrained XLSR embeddings, we apply an information bottleneck that maps the XLSR Encoder output embedding $h$ into a lower-dimensional embedding $h'$. 
The bottleneck we adopt is a two-layer Multi-Layer Perceptron (MLP) with GELU activation.

\subsubsection{Adversarial Training}
\label{sssec:aid_method_5}

Inspired by \cite{webber20_interspeech} in their work of voice anonymisation, we propose training the model to be maximally uncertain about speaker information. This is achieved using a Mean Square Error (MSE) loss $\mathcal{L}_{\text{MSE}}$ between the predicted distribution of speaker labels $p(y_{spk})$ and an even distribution across all speakers $\mathcal{U}(|y_{spk}|)$. The total loss $\mathcal{L}$ is expressed as:
\vspace{-0.7em}
\begin{equation}
    \mathcal{L} = \mathcal{L}_{\text{acc\_clf}} + \alpha \cdot \mathcal{L}_{\text{MSE}}[p(y_{spk}), \mathcal{U}(|y_{spk}|)]\text{,}
    \label{eq:aid_total_loss_v2}
\vspace{-0.7em}
\end{equation}
where $\mathcal{L}_{\text{acc\_clf}}$ is the cross entropy loss for accent classification, and $\alpha$ is a hyperparameter to balance losses.

\vspace{-0.5em}
\subsection{AccentBox: Zero-Shot Accent Generation}
\label{ssec:tts_methods}

\begin{figure}[h!]
\vspace{-1.5em}
\centering
\includegraphics[width=1\linewidth, height=0.88\linewidth, trim = 20 300 15 30, clip]{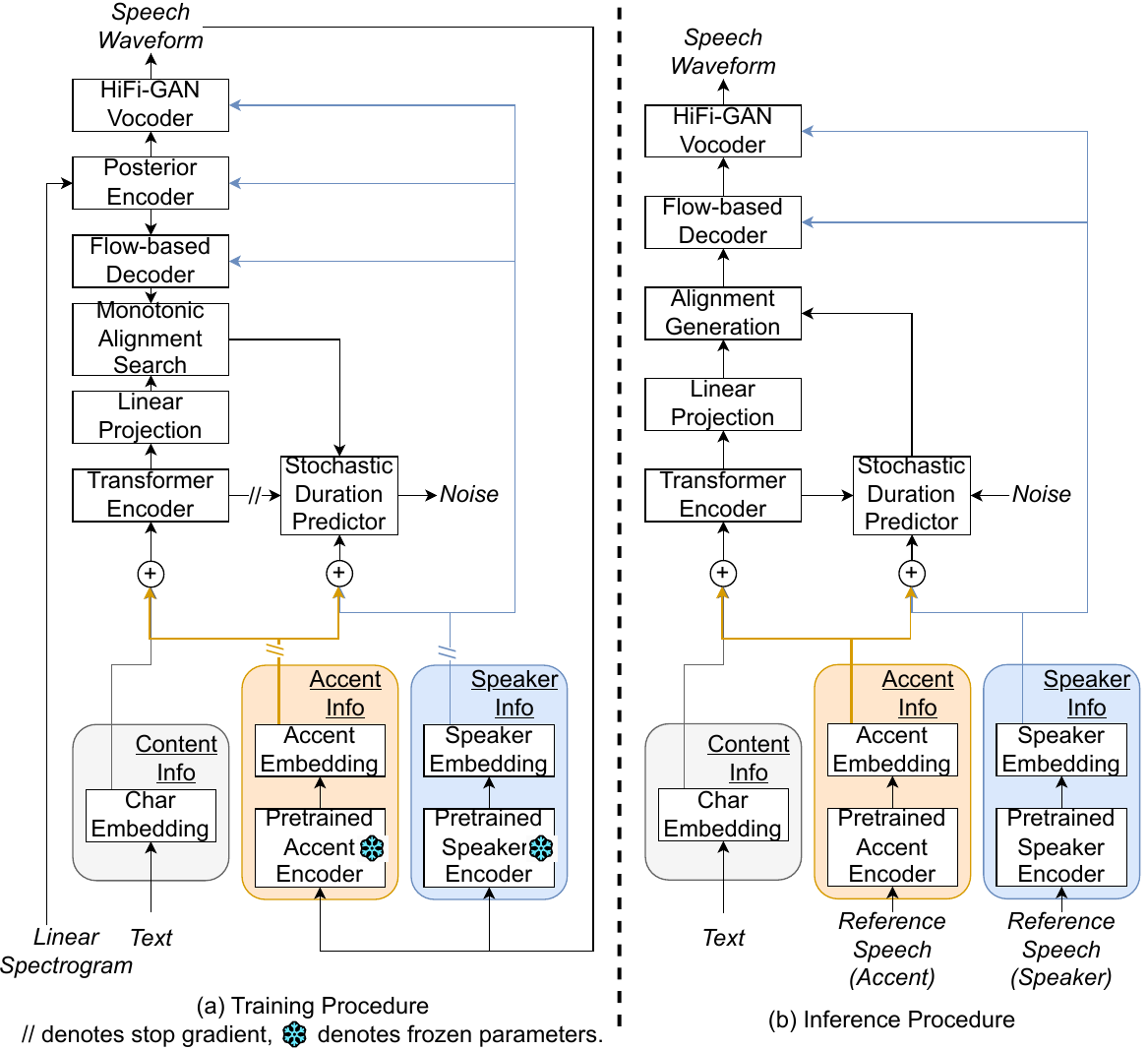}
\captionsetup{justification=centering}
\caption{Model architecture of AccentBox.
\\The pretrained accent encoder (GenAID) is the same as in Fig.\ \ref{fig:aid_methods}.
}
\label{fig:tts_methods}
\vspace{-1.25em}
\end{figure}

In the second stage, we condition a ZS-TTS system on the GenAID embeddings. Fig.\ \ref{fig:tts_methods} shows the model architecture for both training and inference. We build upon YourTTS \cite{pmlr-v162-casanova22a} instead of LLM-based ZS-TTS due to: 1) high data and computation requirements, 2) unstable generation, and 3) lack of open-source models/code. 

\subsubsection{Training}
\label{sssec:tts_methods_train}

Since the same text spoken by speakers of different accents exhibits distinct phonetic and prosodic variations, we condition both the Transformer Text Encoder and the Stochastic Duration Predictor on the accent embeddings learned by GenAID. Compared with YourTTS, we replace the one-hot language embeddings with GenAID accent embeddings as input, as depicted by the pretrained accent encoder (orange block) in Fig.\ \ref{fig:tts_methods}.

\subsubsection{Inference}
\label{sssec:tts_methods_inference}

Table \ref{tab:tts_inference} outlines the different types of inference scenarios explored in this study. All reference speech use unseen speakers as target speaker or accent information, adhering to the zero-shot requirement. \textit{Inherent Accent Generation} examines the hypothesised higher accent fidelity brought by AccentBox; \textit{Cross Accent Generation} examines the hypothesised accent control and disentanglement where speaker and accent conditions mismatch; \textit{Unseen Accent Generation} explores the limits of zero-shot accent generation and tests AccentBox on unseen accents.

\begin{table}[h!]
\vspace{-0.5em}
\centering
\setlength\tabcolsep{2.2pt}
\captionsetup{justification=centering}
\caption{Different types of inference in AccentBox.}
\label{tab:tts_inference}
\vspace{-0.5em}
\begin{tabular}{cccc}
\toprule
Accent Generation & Target Speaker & Target Accent & Speaker-Accent Match? \\ \midrule
Inherent          & Unseen  & Seen   & Yes                   \\
Cross             & Unseen  & Seen   & No                    \\
Unseen            & Unseen  & Unseen & Yes                   \\ \bottomrule
\end{tabular}
\vspace{-1.5em}
\end{table}
\section{Experiments}
\label{sec:exp}

\subsection{GenAID}
\label{ssec:exp_genaid}
\vspace{-0.25em}

\subsubsection{Data}
We make the following modifications to the original CommonAccent processing pipeline to derive a multi-accent speech dataset.
(i) To obtain larger-scale and higher-quality data, we use the latest English portion of Common Voice version 17.0.
(ii) To evaluate the performances of AID models on both seen and unseen speakers, we create separate validation/testing sets for seen and unseen speakers. 
(iii) To train an AID model that generalises well to unseen speakers, we exclude accent labels with insufficient speakers. Most remaining accents have at least 10 speakers with 50 utterances each (for training data and validation/testing on seen speakers), and 20 additional speakers with at least 10 utterances each (for validation/testing on unseen speakers).
(iv) To prevent biasing the AID model towards certain speakers, we allow a maximum of 30 utterances per speaker in the training set. 
Data composition of the final processed data is shown in Appendix I on our demo page.

\subsubsection{Systems}
All five modifications introduced in Sec.\ \ref{ssec:genaid} bring improvement in performance and are accumulatively added (see systems \texttt{\#1}-\texttt{\#6} in Tab.\ \ref{tab:aid_main_results}).

\subsubsection{Configurations}
Following CommonAccent, all systems are initailised from XLSR-large\footnote{\url{https://huggingface.co/facebook/wav2vec2-large-xlsr-53}}. All model parameters are unfrozen in AID finetuning, except for the bottom CNN layers in XLSR Encoder, shown in Fig.\ \ref{fig:aid_methods}. The best system (\texttt{{\#}6}) is trained with a learning rate of 1e-4, bottleneck of 64 dimension, and $\alpha$ of 10.


\subsubsection{Evaluation}
\textit{(i) Classification Metrics:} 
AID performance is evaluated using precision, recall, f1 score, and accuracy. For seen speakers, we report the macro-average across accents to mitigate class imbalance. We also report the f1 score and accuracy gaps between seen and unseen speakers to assess generalisation (smaller gaps indicate better generalisation).
\textit{(ii) T-SNE Visualisation:} 
We visualise speaker and accent information by extracting the latent embeddings before the final classification layer for all utterances in the unseen speaker testing set. These embeddings are processed using t-SNE for visualisation.
\textit{(iii) Silhouette Coefficient for Speaker Clusters (SCSC):} 
To quantify residual speaker information, we group embeddings by speaker for each accent label and calculate the Silhouette coefficient \cite{rousseeuw1987silhouettes}. Lower SCSC values indicate less residual speaker information and more overlap between speaker clusters, as desired.

\vspace{-1em}
\subsection{AccentBox}
\label{ssec:exp_accentbox}
\vspace{-0.25em}

\subsubsection{Systems \& Data}

Table \ref{tab:tts_systems} outlines how different systems are obtained. \texttt{VALL-E X} is the open-source implementation\footnote{\url{https://github.com/Plachtaa/VALL-E-X}}. The \texttt{Pretrained} system is trained on the clean portion of LibriTTS-R \cite{koizumi23_interspeech} for 1 million steps. The remaining three systems are then finetuned on VCTK \cite{vctk} for 200 thousand training steps with different accent conditioning. 
11 speakers (one for each accent) are reserved for inference only (including 9 seen and 2 unseen accents). To test the performances of different systems in terms of accent generation, we use an elicitation passage of 23 sentences, \textit{Comma Gets a Cure}\footnote{\url{https://www.dialectsarchive.com/CommaGetsACure.pdf}}, as input text, and a fixed utterance (24th utterance from each speaker) as reference speech, to avoid the influence of reference speech content.

\begin{table}[h!]
\vspace{-0.5em}
\centering
\setlength\tabcolsep{1.5pt}
\captionsetup{justification=centering}
\caption{Comparison of different ZS-TTS systems.}
\label{tab:tts_systems}
\vspace{-0.5em}
\begin{tabular}{cccc}
\toprule
System              & Data             & Accent Info       & Initialisation         \\ \midrule
\texttt{VALL-E X} & Unknown             & N/A               & inference only \\ \midrule
\texttt{Pretrained} & LibriTTS-R clean             & N/A               & from scratch           \\ 
\texttt{Baseline}   & VCTK & N/A               & from \texttt{Pretrained} \\
\texttt{Accent\_ID} & VCTK             & one-hot embedding & from \texttt{Pretrained} \\
\texttt{Proposed}   & VCTK             & GenAID embedding  & from \texttt{Pretrained} \\ \bottomrule
\end{tabular}
\vspace{-0.5em}
\end{table}

\subsubsection{Configurations}

To ensure high audio quality in synthesis, all waveforms are downsampled to 24 kHz as target waveform. We train all models with a batch size of 32 and an initial learning rate of 2e-4.

\subsubsection{Objective Evaluation}

\textit{(i) Accent Cosine Similarity (AccCos):} We use two AID models \texttt{{\#}4} and \texttt{{\#}6} to extract accent embeddings, and calculate cosine distances between reference and generated speech, avoiding biases towards \texttt{Proposed} which is conditioned on embeddings from \texttt{{\#}6}.
\textit{(ii) Speaker Cosine Similarity (SpkCos):} We use Resemblyzer\footnote{\url{https://github.com/resemble-ai/Resemblyzer}} \cite{8462665} to extract speaker embeddings of generated speech and compare them to reference speech (speaker) for cosine distance calculation.
\textit{(iii) Why no Word Error Rate (WER)?} As verified by \cite{10095057}, various SOTA ASR models have clear bias against accents and WER varies across different accents. A high WER could indicate either unclear or more accented generation which makes ASR models harder to recognise correctly.

\subsubsection{Subjective Evaluation}

To holistically evaluate different aspects of generated speech, we ask listeners to compare different systems based on three metrics: i) \textit{accent similarity}, ii) \textit{speaker similarity}, and iii) \textit{naturalness}.
To fully compare all systems, we conduct ABC ranking tests (\texttt{Baseline} vs \texttt{Accent\_ID} vs \texttt{Proposed}) for inherent accent generation and AB preference tests (\texttt{Accent\_ID} vs \texttt{Proposed}) for cross accent generation. The \texttt{Baseline} does not take any accent information as input condition and cannot perform cross accent generation, therefore it is not evaluated in the latter task. 
All listeners are recruited through Prolific\footnote{\url{https://www.prolific.com}} from target accent regions. Ten listeners are recruited for each utterance.
Due to budget constraints, we are only able to conduct listening tests on two accents. We choose American and Irish accents, with different data size (8.03 and 3.03 hours respectively) in the finetuning data.

\section{Results and Analysis}
\label{sec:results}

\subsection{GenAID}
\label{ssec:results_genaid}
\vspace{-0.25em}

\begin{figure}[h!]
     \centering
     \begin{subfigure}[h!]{0.24\textwidth}
         \centering
         \includegraphics[width=\textwidth, trim = 0 12 135 0, clip]{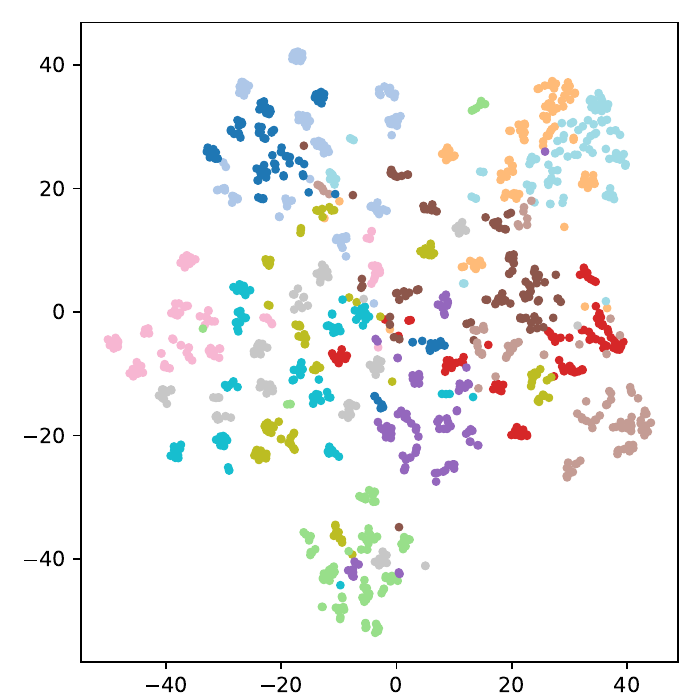}
         \caption{{\#}1 baseline}
         \label{fig:aid_tsne_E5}
     \end{subfigure}
     \hfill
     \begin{subfigure}[h!]{0.24\textwidth}
         \centering
         \includegraphics[width=\textwidth, trim = 0 12 135 0, clip]{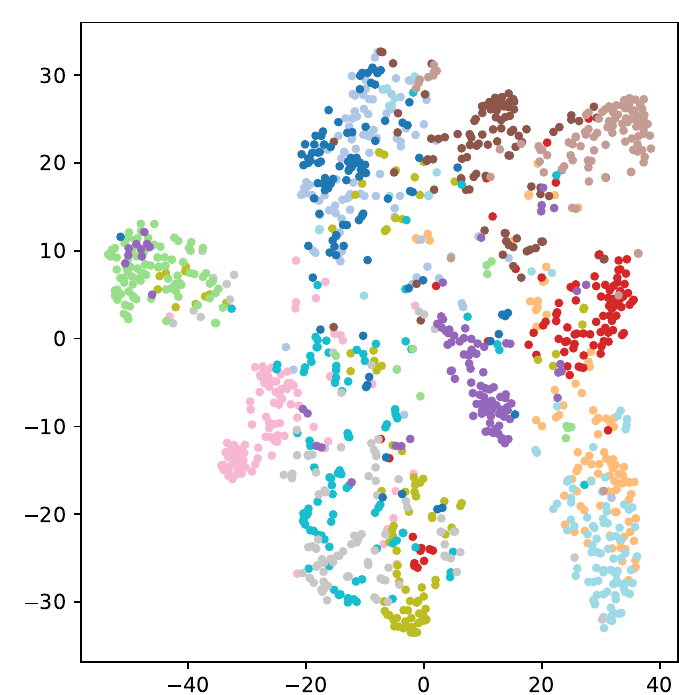}
         \caption{{\#}6 w/ adv.\ training}
         \label{fig:aid_tsne_X6}
     \end{subfigure}
        \vspace{-0.5em}\captionsetup{justification=centering}
        \caption{T-SNE visualisation of embeddings by different AID systems on unseen speakers. (Each color represents an accent.)}
        \label{fig:aid_tsne}
\vspace{-1.5em}
\end{figure}

Table \ref{tab:aid_main_results} shows the AID resulst of different systems. On \textbf{unseen speakers}, which we focus on, a significant 0.15 f1 score and 0.13 accuracy improvement is achieved (\texttt{{\#}1} vs \texttt{{\#}6}). The best system (\texttt{{\#}6}) achieves a \textbf{0.56} AID accuracy on unseen speakers, significantly better than the 0.08 random baseline. We also reduced speaker entanglement, with smaller accuracy gaps between seen and unseen speakers (0.53 vs 0.06 by \texttt{{\#}1} vs \texttt{{\#}6}), and lower SCSC (0.236 vs 0.079 by \texttt{{\#}1} vs \texttt{{\#}6}). Note that high accuracy on seen speakers with a large gap to unseen speakers is not desirable, as this suggests the model is memorizing speaker-accent mappings rather than learning to discriminate accents. Of all the proposed modifications, we find information bottleneck to be the most effective. We further visualise embeddings of \texttt{{\#}1} and \texttt{{\#}6} on unseen speakers using t-SNE. The best system (\texttt{{\#}6}) shows better-separated accent clusters and less speaker-accent entanglement compared to the baseline (\texttt{{\#}1}) in Fig.\ \ref{fig:aid_tsne}.


\begin{table}[h!]
\vspace{-0.5em}
\centering
\setlength\tabcolsep{1.5pt}
\captionsetup{justification=centering}
\caption{Accent identification results. All ``w/" changes are \underline{accumulative}. ``adv." - adversarial; ``prec" - precision; ``rec" - recall.}
\label{tab:aid_main_results}
\vspace{-0.5em}
\begin{tabular}{l|cc|cccc|cc|c}
\toprule
\multicolumn{1}{c|}{\multirow{2}{*}{AID Systems}} & \multicolumn{2}{c|}{Seen Spks} & \multicolumn{4}{c|}{Unseen Spks$\uparrow$} & \multicolumn{2}{c|}{Gap$\downarrow$} & \multicolumn{1}{c}{\multirow{2}{*}{SCSC$\downarrow$}}\\ \cmidrule{2-9} 
\multicolumn{1}{c|}{}       & f1   & acc  & prec & rec  & f1   & acc  & f1   & acc  \\ \midrule
\#1 baseline & 0.95 & 0.96 & 0.56 & 0.43 & 0.40 & 0.43 & 0.55 & 0.53 & 0.236 \\
\#2 w/ valid on unseen & 0.82 & 0.86 & 0.57 & 0.47 & 0.45 & 0.47 & 0.37 & 0.39 & 0.142 \\
\#3 w/ weighted sampler & 0.77 & 0.58 & 0.56 & 0.47 & 0.46 & 0.47 & 0.31 & 0.11 & 0.167 \\
\#4 w/ perturbation & 0.81 & 0.63 & 0.60 & 0.50 & 0.48 & 0.50 & 0.33 & 0.13 & 0.176 \\
\#5 w/ bottleneck & 0.73 & 0.66 & 0.61 & \textbf{0.56} & \textbf{0.55} & \textbf{0.56} & \textbf{0.18} & 0.10 & 0.090  \\
\#6 w/ adv.\ training & 0.78 & 0.62 & \textbf{0.63} & \textbf{0.56} & \textbf{0.55} & \textbf{0.56} & 0.23 & \textbf{0.06} & \textbf{0.079} \\ \bottomrule
\end{tabular}
\vspace{-2em}
\end{table}


\subsection{AccentBox}
\label{ssec:results_accentbox}
\vspace{-0.25em}

Table \ref{tab:tts_obj_results} shows the \underline{objective} evaluation results of all 5 systems. Table \ref{tab:tts_subj_results_inh} shows the subjective evaluation results for \textit{Inherent Accent Generation} by comparing the preferences among the three finetuned systems. Table \ref{tab:tts_subj_results_cross} shows the subjective evaluation results for \textit{Cross Accent Generation} by comparing the preferences between the only two systems which can perform accent conversion.

For the task of \textit{Unseen Accent Generation} which is significantly more difficult, requiring TTS models to generalise to unseen accents, we include generated audios on the demo page with comparison between \texttt{Baseline} and \texttt{Proposed}. We leave more systematic evaluation of such a task for future work.

\subsubsection{Inherent Accent Generation}
\label{sssec:tts_results_inherent}
\vspace{-0.25em}

The \texttt{Proposed} system achieves higher accent similarity across both objective and subjective evaluations. It outperforms other systems, including the open-source \texttt{VALL-E X}, in objective evaluations, regardless of the model used to extract accent embeddings. In subjective evaluations, the \texttt{Proposed} system consistently surpasses both the \texttt{Baseline} and \texttt{Accent\_ID} systems in generating American and Irish accents, demonstrating superior accent fidelity for inherent accent generation. For speaker similarity, while subjective evaluations favour the \texttt{Proposed} system, objective evaluations show lower speaker cosine similarity scores, likely due to bias in the speaker verification model towards common accents or listeners’ difficulty in distinguishing accent and speaker identity. The \texttt{Proposed} system also shows higher naturalness when generating American accents but receives lower preference for Irish accents, potentially due to limited Irish accent data and the system’s sensitivity to monotonic prosody in reference speech. Further research with larger, more diverse datasets and refined evaluation methods is needed to better understand these discrepancies and improve performance across accents.

\begin{table}[h!]
\centering
\setlength\tabcolsep{2pt}
\captionsetup{justification=centering}
\caption{\underline{Objective} evaluation on 9 seen accents.
\\AccCos - Accent Cosine Similarity, SpkCos - Speaker Cosine Similarity. \texttt{{\#}4} and \texttt{{\#}6} are two AID systems in Table \ref{tab:aid_main_results}.
}
\label{tab:tts_obj_results}
\vspace{-0.5em}
\begin{tabular}{c|ccc|ccc}
\toprule
\multirow{2}{*}{System} & \multicolumn{3}{c|}{Inherent Accent Generation}      & \multicolumn{3}{c}{Cross Accent Generation}         \\ \cmidrule{2-7} 
 &
  \begin{tabular}[c]{@{}c@{}}AccCos\\ (\texttt{{\#}4})\end{tabular} &
  \begin{tabular}[c]{@{}c@{}}AccCos\\ (\texttt{{\#}6})\end{tabular} &
  SpkCos &
  \begin{tabular}[c]{@{}c@{}}AccCos\\ (\texttt{{\#}4})\end{tabular} &
  \begin{tabular}[c]{@{}c@{}}AccCos\\ (\texttt{{\#}6})\end{tabular} &
  SpkCos \\ \midrule
\texttt{VALL-E X}              & 0.7801          & 0.9077          & \textbf{0.8605} & /               & /               & /               \\ \midrule
\texttt{Pretrained}              & 0.7510          & 0.8911          & 0.8413          & /               & /               & /               \\
\texttt{Baseline}                & 0.7232          & 0.8989          & 0.8362          & /               & /               & /               \\
\texttt{Accent\_ID}              & 0.7837          & 0.9291          & 0.8386          & 0.7350          & 0.8985          & 0.8073          \\
\texttt{Proposed}                & \textbf{0.8037} & \textbf{0.9336} & 0.8293          & \textbf{0.7538} & \textbf{0.9067} & \textbf{0.8100} \\ \bottomrule
\end{tabular}
\end{table}

\begin{table}[h!]
\vspace{-0.5em}
\centering
\setlength\tabcolsep{1pt}
\captionsetup{justification=centering}
\caption{Subjective evaluation for \textit{Inherent Accent Generation}.
\\``Sim." - similarity. ``Pref." - preference rate for \texttt{Proposed}.
\\{*}: weak statistical significance.}
\label{tab:tts_subj_results_inh}
\vspace{-0.25em}
\begin{tabular}{cccccccc}
\toprule
\multirow{2}{*}{Comparison} & \multirow{2}{*}{Accent} & \multicolumn{2}{c}{Accent Sim.} & \multicolumn{2}{c}{Speaker Sim.} & \multicolumn{2}{c}{Naturalness} \\ \cmidrule{3-8} 
                              &          & Pref. & p-value   & Pref. & p-value   & Pref. & p-value  \\ \midrule
\multirow{2}{*}{vs \texttt{Baseline}}  & US & \textbf{69.1\%}     & 1.8E-04  & \textbf{70.0\%}     & 1.2E-03  & \textbf{60.0\%}     & 1.1E-02 \\
                              & Irish    & \textbf{61.3\%}     & 1.4E-02  & \textbf{57.8\%}     & 9.4E-02* & \textit{33.9\%}   & 2.8E-03 \\ 
\multirow{2}{*}{vs \texttt{Accent\_ID}} & US & \textbf{57.4\%}     & 8.4E-02* & \textbf{62.2\%}     & 2.1E-02  & \textbf{56.1\%}     & 3.4E-02 \\
                              & Irish    & \textbf{65.7\%}     & 4.9E-06  & \textbf{59.1\%}     & 9.3E-03  & \textit{43.9\%}   & 2.6E-02 \\ \bottomrule
\end{tabular}
\end{table}

\begin{table}[h!]
\centering
\setlength\tabcolsep{1pt}
\captionsetup{justification=centering}
\caption{Subjective evaluation for \textit{Cross Accent Generation}. 
}
\label{tab:tts_subj_results_cross}
\vspace{-0.25em}
\begin{tabular}{cccccccc}
\toprule
\multirow{2}{*}{Comparison} & \multirow{2}{*}{Accent} & \multicolumn{2}{c}{Accent Sim.} & \multicolumn{2}{c}{Speaker Sim.} & \multicolumn{2}{c}{Naturalness} \\ \cmidrule{3-8} 
                              &          & Pref. & p-value  & Pref. & p-value  & Pref. & p-value  \\ \midrule
\multirow{2}{*}{vs \texttt{Accent\_ID}} & US & \textbf{70.0\%}     & 1.1E-06 & \textit{45.2\%}   & 3.2E-02 & \textbf{65.2\%}     & 1.5E-04 \\
                              & Irish    & \textbf{61.7\%}     & 1.3E-02 & \textbf{61.3\%}     & 1.1E-02 & \textbf{63.0\%}     & 3.1E-02 \\ \bottomrule
\end{tabular}
\vspace{-1.5em}
\end{table}

\subsubsection{Cross Accent Generation}
\label{sssec:tts_results_cross}

The overall objective results for cross-accent generation are lower than those for inherent accent generation, indicating that accent conversion is a more challenging task. The \texttt{Proposed} system demonstrates higher accent similarity in both objective and subjective evaluations, showing superior accent fidelity in accent conversion. However, subjective speaker similarity results are mixed, with higher preference for Irish but not for American accents. This may stem from listeners perceiving generated speech with higher accent similarity as more distinct in speaker identity from the original English-accented reference speech. In terms of naturalness, the \texttt{Proposed} system outperforms, likely due to more consistent accent generation during conversion. In contrast, the \texttt{Accent\_ID} system, which relies on one-hot accent labels from limited TTS data, struggles with accent consistency, as swapping one-hot accent embeddings forces the model to generalise to unseen speaker-accent pairs with insufficient information.

\vspace{-0.5em}
\section{Conclusions}
\vspace{-0.5em}

In this work, we introduce zero-shot accent generation and a novel two-stage pipeline as a benchmark. In the first stage, AID, we verify, quantify, and address speaker-accent entanglement, with SOTA performance of 0.56 f1 score in 13-accent classification on unseen speakers. In the second stage, zero-shot accent generation, we highlight and address the problem of accent mismatch/hallucination in ZS-TTS, with better accent fidelity in inherent/cross accent generation while enabling unseen accent generation. In the future, we will further experiment with L2 accent generation/conversion.
\vspace{-0.5em}
\section*{Acknowledgement}
\vspace{-0.5em}
This work was supported in part by the UKRI AI Centre for Doctoral Training (CDT) in Responsible and Trustworthy in-the-world NLP (Grant EP/Y030656/1), School of Informatics, and School of Philosophy, Psychology \& Language Sciences, the University of Edinburgh.

\clearpage
\newpage
\bibliographystyle{IEEEtran}
\bibliography{mybibfile}

\end{document}